\documentclass[prl,twocolumn,superscriptaddress]{revtex4}

\usepackage{amsmath,amsfonts,amssymb,amstext}
\usepackage{graphicx}
\usepackage{times}
\usepackage{bbm}
\usepackage{color}
\usepackage[colorlinks,linkcolor=blue]{hyperref}

\def\>{\rangle}
\def\<{\langle}
\def\({\left(}
\def\){\right)}

\newcommand{\bbraket}[2]{\mbox{$\big(#1  , #2 \big)$}} 

\newcommand{\ket}[1]{|#1\>}
\newcommand{\bra}[1]{\<#1|}

\DeclareMathOperator{\Tr}{Tr}
\DeclareMathOperator{\tr}{tr}

\DeclareMathOperator{\plog}{polylog}
\DeclareMathOperator{\sign}{sign}

\DeclareMathOperator{\range}{range}
\DeclareMathOperator{\Span}{span}

\def\CC{\mathbbm{C}}

\def\EE{\mathbbm{E}}

\def\P{\mathcal{P}}
\def\R{\mathcal{R}}
\def\Q{\mathcal{Q}}
\def\A{\mathcal{A}}

\def\Id{\mathbbm{1}}

\newcommand{\set}[1]{\lbrace #1 \rbrace}
\newcommand{\norm}[1]{\lVert #1 \rVert}

\newtheorem{theorem}{Theorem}

\newtheorem{observation}{Observation}

\begin{document}

\title{Quantum state tomography via compressed sensing}

\author{David Gross}
\affiliation{Institute for Theoretical Physics, Leibniz University Hannover, 30167 Hannover, Germany}

\author{Yi-Kai Liu}
\affiliation{Institute for Quantum Information, California Institute of Technology, Pasadena, CA, USA}

\author{Steven T. Flammia}
\affiliation{Perimeter Institute for Theoretical Physics, Waterloo, Ontario, N2L 2Y5 Canada}

\author{Stephen Becker}
\affiliation{Applied and Computational Mathematics, California Institute of Technology, Pasadena, CA, USA}

\author{Jens Eisert}
\affiliation{Institute of Physics und Astronomy, University of Potsdam, 14476 Potsdam, Germany}

\date{July 11, 2010}

\begin{abstract}
We establish methods for quantum state tomography based on
compressed sensing. These methods are specialized for quantum states that are
fairly pure, and they offer a significant performance improvement on large
quantum systems. In particular, they are able to reconstruct an unknown
density matrix of dimension $d$ and rank $r$ using $O(rd \log^2 d)$ measurement
settings, compared to standard methods that require $d^2$ settings. Our methods
have several features that make them amenable to experimental implementation: 
they require only simple Pauli measurements, use fast convex optimization, 
are stable against noise, and can be applied to states that are only approximately 
low-rank. The acquired data can be used to certify that the state is indeed 
close to pure, so no {\it a priori\/} assumptions are needed.  
We present both theoretical bounds and numerical simulations.
\end{abstract}

\maketitle

The tasks of reconstructing the quantum states and processes produced
by physical systems --- known respectively as quantum state and process
tomography~\cite{Paris2004} --- are of increasing importance in physics
and especially in quantum information science.  Tomography has been
used to characterize the quantum state of trapped
ions~\cite{Haffner2005} and an optical entangling
gate~\cite{OBrien2003} among many other implementations.  But a
fundamental difficulty in performing tomography on many-body systems is
the exponential growth in the state space dimension.
For example, to get a maximum-likelihood estimate of a quantum state
of $8$ ions, Ref.~\cite{Haffner2005} required 
hundreds of thousands of measurements and weeks of post-processing.

Still, one might hope to overcome this obstacle, because the vast
majority of quantum states are not of physical interest.  Rather, one
is often interested in states with special properties: pure
states, states with particular symmetries, ground states of local
Hamiltonians, etc., and tomography might be more efficient in such special cases~\cite{Kaznady2009}. 

In particular, consider pure or nearly pure quantum states, i.e., states with low entropy. 
More precisely,
consider a quantum state that is essentially supported on an
$r$-dimensional space, meaning the density matrix is close 
(in a given norm) 
to a matrix of rank $r$, where $r$ is small.  Such states arise in
very common physical settings, e.g.\ a pure state subject to a
local noise process \cite{localnoise}.

A standard implementation of tomography \cite{S1,S2} would use $d^2$
or more measurement settings, where $d=2^n$ for an $n$-qubit system.
But a simple parameter counting argument suggests that $O(rd)$
settings could possibly suffice --- a significant improvement.
However, it is not clear how to achieve this performance in practice,
i.e., how to choose these measurements, or how to efficiently
reconstruct the density matrix.  For instance, the problem of finding a
minimum-rank matrix subject to linear constraints is NP-hard in
general~\cite{natarajan95}.

In addition to a reduction in experimental complexity, one might hope
that a post-processing algorithm which takes as input only $O(rd)\ll
d^2$ numbers could be tuned to run considerably faster than standard
methods. Since the output of the procedure is a 
low-rank approximation to the density operator and only requires
$O(rd)$ numbers be specified, it becomes conceivable that the
run time scales better than $O(d^2)$, clearly impossible for naive
approaches using dense matrices. 

In this Letter, we introduce a method to achieve such drastic
reductions in measurement complexity,
together with efficient algorithms for post-processing.
The approach further develops ideas
that have recently been studied under the label of
``compressed sensing''. Compressed sensing
\cite{compressedsensing} provides techniques for
recovering a sparse vector from a small number of measurements~\cite{Kosut2008}.
Here, sparsity means that this vector contains only a few non-zero entries
in a specified basis, and the measurements are linear functions of its
entries.  When the measurements are chosen at random (in a certain
precise sense), then with high probability two surprising things
happen:  the vector is uniquely determined by a small number of
measurements, and it can be recovered by an efficient
convex optimization algorithm~\cite{compressedsensing}.

Matrix completion~\cite{Candes2008,Candes2009a, Candes2009} is a
generalization of compressed sensing from vectors to matrices.  Here,
one recovers certain ``incoherent'' low-rank matrices $X$ from a small
number of matrix elements $X_{i,j}$.  The problem of low-rank quantum
state tomography bears a strong resemblance to matrix completion.
However, there are important differences.  We wish to use
measurements that can be more easily implemented in an experiment than
obtaining elements $\rho_{i,j}$ of density matrices.  Previous results
\cite{Candes2008,Candes2009a, Candes2009} cannot be applied to this
more general situation.  We would also like to avoid any unnatural
incoherence assumptions crucial in prior work \cite{Candes2008}. 

Our first result is a protocol for tomography that
overcomes both of these difficulties:  it uses Pauli measurements only, 
and it works for arbitrary density matrices.  
We prove that only $O(r d \,\log^2 d)$ measurement settings suffice.
What is more, our proof introduces some new techniques, which both
generalize and vastly simplify the previous work on matrix completion.
We sketch the proof here; a more complete version appears in
\cite{prep2}.  This provides the basic theoretical justification for
our method of doing tomography.

We then consider a number of practical issues.  In a real experiment,
the measurements are noisy, and the true state is only approximately
low-rank.  We show that our method is robust to these sources of
error.  
We also describe ways to certify that a state is nearly pure
without any \textit{a priori} assumptions.

Finally, we present fast algorithms for reconstructing the density
matrix from the measurement statistics based on semidefinite
programming -- a feature not present in earlier methods for
pure-state tomography \cite{Kaznady2009, S1,S2}.  These are adapted from algorithms for matrix
completion \cite{Cai2008}, and they are much faster than standard
interior-point solvers.  Reconstructing a low-rank density matrix for
$8$ qubits takes about one minute on an ordinary laptop computer. 


While our methods do not overcome the exponential growth in
measurement complexity (which is provably impossible for any protocol
capable of handling generic pure states), they do significantly push the boundary of
what can be done in a realistic setting. 

Our techniques also apply to process tomography:  to
characterize an unknown quantum process $\mathcal{E}$, prepare
the Jamio{\l}kowski state $\rho_{\mathcal{E}}$, and perform state
tomography on $\rho_{\mathcal{E}}$. Our methods work when
$\mathcal{E}$ can approximately be written as a sum of only a few
Kraus operators, because this implies that $\rho_{\mathcal{E}}$ has
small rank.


\paragraph{Matrix recovery using Pauli measurements.}  
We consider the case of $n$ spin-$1/2$ systems in an unknown state
$\rho$ 
\cite{note1}. An $n$-qubit Pauli matrix is of the form $w=\bigotimes_{i=1}^n w_i$,
where $w_i \in \set{\mathbbm{1}, \sigma^x, \sigma^y, \sigma^z}$. There
are $d^2$ such matrices, labeled $w(a), a\in[1,d^2]$. The protocol
proceeds as follows: choose $m$ integers $A_1, \dots, A_m \in [1,d^2]$
at random and measure the expectation values $\tr\rho w(A_i)$. 
One then solves a convex optimization problem:
minimize $\|\sigma\|_{\tr}$ 
\cite{note2} subject to 
\begin{equation}\label{eqn:feasible}
	\tr\sigma = 1, \qquad 
	\tr w(A_i) \sigma = \tr w(A_i)
	\rho.
\end{equation}

\begin{theorem}[Low-rank tomography]\label{lr}
Let $\rho$ be an arbitrary state of rank $r$.  If $m=c d r \log^2d$
randomly chosen Pauli expectations are known, then $\rho$ can be
uniquely reconstructed by solving the convex optimization problem
(\ref{eqn:feasible}) with probability of failure exponentially small
in $c$.
\end{theorem}

The proof is inspired by, but technically very different from, earlier
work on matrix completion \cite{Candes2008}.  
Our methods are more general, can
be tuned to give tighter bounds, and are much more compact, allowing
us to present a fairly complete argument in this Letter. A more
detailed presentation of this technique -- covering the reconstruction
of low-rank matrices from few expansion coefficients w.r.t.\ general
operator bases (not just Pauli matrices or matrix elements) -- will be
published elsewhere \cite{prep2}.


{\it Proof:} 
Here we sketch the argument and explain the main ideas; detailed calculations are in the EPAPS supplement. 

Note that the linear constraints (\ref{eqn:feasible}) depend only on
the projection of $\rho$ onto the span of the measured observables
$w(A_1),\dots,w(A_m)$. This is precisely the range of the ``sampling
operator'' $\R: \rho
\mapsto  \frac{d}{m}\sum_{i=1}^m w(A_i) \tr \rho w(A_i)$.
(Note that $\EE[\R(\rho)] = \rho$.) 
Indeed, the convex program can be
written as $\min_\sigma \|\sigma\|_{\tr}$ s.t.\ $\R\sigma=\R\rho$.
Evidently, the solution is unique if for all deviations
$\Delta:=\sigma-\rho$ away from $\rho$ either $\R\Delta \neq 0$ or
$\|\rho+\Delta\|_{\tr} > \|\rho\|_{\tr}$.  

We will ascertain this by using a basic idea from convex optimization:
constructing a \emph{strict subgradient} $Y$ for the norm.  A
matrix $Y$ is a strict subgradient if $\|\rho+\Delta\|_{\tr} > \|\rho\|_{\tr}
+ \tr Y\Delta$ for all $\Delta \neq 0$. The main contribution below is
a method for constructing such a $Y$ which is also in the range of
$\R$. For then $\R\Delta=0$ implies that $\Delta$ is orthogonal to the
range of $\R$, thus $\tr Y\Delta =0$ and the subgradient condition
reads $\|\rho+\Delta\|_{\tr} > \|\rho\|_{\tr}$.  This implies
uniqueness. (In fact, it is sufficient to
approximate the subgradient condition in a certain sense).

Let $E$ be the projection onto the range of $\rho$, let $T$ be the
space spanned by those operators whose row or column space is contained in
$\range\rho$.  
Let $\P_T$ be the projection
onto $T$, $\P_T^\bot$ onto the orthogonal complement. Decompose
$\Delta = \Delta_T + \Delta_T^\bot$, the parts of $\Delta$ that lie in
the subspaces $T$ and $T^\bot$. We distinguish two cases:
\textit{(i)} $\|\Delta_T\|_2 > d^2\|\Delta_T^\bot\|_2$, and
\textit{(ii)} $\|\Delta_T\|_2 \leq d^2\|\Delta_T^\bot\|_2$
\cite{note2}.

Case \textit{(i)} is easier. In this case, $\Delta$ is
well-approximated by $\Delta_T$ and essentially we only have to show
that the restriction $\A:=\P_T\R\P_T$ of $\R$ to $T$ is invertible.
Using a non-commutative large deviation bound (see EPAPS supplement),
\begin{equation}\label{eqn:devNorm}
	\Pr[\|\A-\Id_T\| > t]< 4dr e^{-t^2\kappa /8}
\end{equation}
where $\kappa=m/(dr)$ \cite{note2}. Hence the probability that
$\|\A-\Id_T\|>\frac12$ is smaller than $4dre^{-\kappa/32}=:p_1$.
If that is not the case, one easily sees that $\|\R\Delta\|_2 > 0$,
concluding the proof for this case.

Case \textit{(ii)} is more involved. A matrix $Y
\in\Span(w(A_1),\ldots,w(A_m))$ is an \emph{almost subgradient} 
\cite{note3} if
\begin{equation}\label{eqn:dual}
	\|\P_T Y - E\|_2 \leq 1/(2 d^2), \qquad \|\P_T^\bot Y \| < 1/2.
\end{equation}
First, suppose such a $Y$ exists. Then a simple calculation (see EPAPS)
using the condition \textit{(ii)} shows that $\R\Delta=0$ indeed implies
$\|\rho+\Delta\|_{\tr} > 
\|\rho\|_{\tr}$ as hinted at
above. This proves uniqueness in case \textit{(ii)}.
The difficult part consists in showing that an almost-subgradient
exists. 

To this end, we
design a recursive process (the ``golfing scheme'' \cite{prep2}) which
converges to a subgradient exponentially fast.
Assume we draw $l$ batches of $\kappa_0 r d$ Pauli observables
independently at random ($\kappa_0$ will be chosen later).
Define recursively $X_0=E$, 
\begin{equation}
Y_i=\sum_{j=1}^i \R_j X_{j-1}, \quad X_i = E - \P_T Y_i, 
\end{equation}
$Y = Y_l$. Let $\R_i$ be the sampling operator associated with the
$i$th batch, and $\A_i$ its restriction to $T$.  Assume that in each
run $\|\A_i-\Id_T\|_2<1/2$. Denote the probability of this event not
occurring by $p_2$. Then
\begin{eqnarray*}
	\|X_i\|_2 
	&=&
	\|X_{i-1}-\P_T \R_i X_{i-1}\|_2 \\
	&=& \|(\Id_T - \A_i) X_{i-1}\|_2 
	\leq 1/2 \|X_{i-1}\|_2,
\end{eqnarray*}
so that $\|X_i\|_2\leq 2^{-i} \|X_0\| = 2^{-i} \sqrt r$.
Hence, $Y=Y_l$ fulfills the first part of (\ref{eqn:dual}), as soon as
$l\geq \log_2 (2 d^2 \sqrt r)$. We turn to the second part.
Again using large-deviation techniques (EPAPS) we
find 
$\|\P_T^\bot\R_i X_{i-1}\| \leq 1/(4\sqrt r) \|X_{i-1}\|_2$ with
some (high) probability $(1-p_3)$. Therefore:
\begin{eqnarray}\label{eqn:golfing}
	\|\P_T^\bot Y_l\| 
	\leq \sum_{j=1}^l \|\P_T^\bot \R_j X_{j-1}\| 
	\leq \frac14 \sum_{j=0}^\infty 2^{-l} < \frac12,
\end{eqnarray}
which is the second part of (\ref{eqn:dual}).

Lastly, we have to bound the total probability of failure $p_f \leq p_1 +
p_2 +p_3$. Set $\kappa_0 = 64 \mu (1+\ln(8 dl))$,
which means that $m=d r (\ln d)^2\,O(1)$ coefficients will be sampled
in total. A simple calculation gives 
$p_f \leq e^{-\mu}$.
This completes the proof of our main result. $\Box$

In the remaining space, we address the important aspects of resilience
against noise, certified tomography, and numerical performance. Owing to space
limitations, the presentation will focus on conceptual issues, with the details in~\cite{prep1}.

\paragraph{Robustness to noise.} Realistic situations will differ from
the previous case in two regards. First, the true state $\rho_t$ may
not be low-rank, but only well approximated by a state $\rho$ of rank
$r$: $\|\rho_t - \rho\|_2 \leq \varepsilon_1$.
Second, due to systematic and statistical noise, the available
estimates for the Pauli expectations are not exactly $\tr \rho_t w(a)$,
but of the form $\tr \omega w(a)$ for some 
matrix $\omega$.  Assume $\|\R \omega - \R\rho_t\|_2\leq \varepsilon_2$
(in practical situations, $\varepsilon_2$ may be estimated from the error
bars associated with the individual Pauli expectation values
\cite{errormodel}).
In order to get an estimate for $\rho_t$, choose some $\lambda \geq 1$ and 
$\varepsilon \geq \lambda (\sqrt{d^2/m}) \varepsilon_1 + \varepsilon_2$, 
and solve the convex program 
\begin{equation}\label{E:cop}
	\min  \|\sigma\|_{\tr} \ , \ \text{subject to}\,  
	\|  \R\sigma- \R\omega\|_2 \leq \varepsilon.
\end{equation}

\begin{observation}[Robustness to noise]\label{robust}
	Let $\rho_t$ be an approximately low-rank state as described above. 
	Suppose $m=c d r \log^2d$ randomly chosen Pauli expectations are known 
	up to an error of $\varepsilon$ as in (\ref{E:cop}), 
	and let $\sigma^\star$ be the solution of (\ref{E:cop}).
	Then the difference $\|\sigma^\star - \rho_t\|_{\tr}$ is
	smaller than $O(\varepsilon \sqrt{rd})$. This holds with probability of 
	failure at most $1/\lambda^2$ plus the probability of failure in 
	Theorem~1.
\end{observation}

The proof combines ideas from Ref.\
\cite{Candes2009} with our argument above 
\cite{note4}.
The main difference from the noise-free case 
is that, instead of using $\tr Y\Delta=0$, we must
now work with $|\tr Y\Delta| \leq 2\|Y\|_2\,\delta$. With this estimate,
Observation 1 follows from the noise-free proof, together with some
elementary calculations (see EPAPS).  We remark that the above
bound is likely to be quite loose; based on related work involving the
``restricted isometry property,'' we conjecture that the robustness to
noise is actually substantially stronger than what is shown here
\cite{FCRP08}.

\paragraph{Certified tomography of almost pure states.} The preceding
results require an {\it a priori\/} promise: that the true state
$\rho_t$ is $\delta_1$-close to a rank-$r$ state.  However, when
performing tomography of an unknown state, neither $r$ nor $\delta_1$
are known beforehand.  There are a few solutions to this quandary.
First, $r$ and $\delta_1$ may be estimated from other physical
parameters of the system, such as the strength of the local noise
\cite{localnoise}.

Another approach is to estimate $r$ and $\delta_1$ from the same data
that is used to reconstruct the state.  When $r=1$, this approach is
particularly effective, in entirely assumption-free tomography:  
one can estimate $\delta_1$, using only
$O(d)$ Pauli expectation values.  This is because $\delta_1$ is
related to the purity $\Tr \rho^2$, which has a simple closed-form
expression in terms of Pauli expectation values.  See EPAPS for
details.  We get:

\begin{observation}[Certified tomography]
	Assume that the unknown physical state is close to being pure. Then
	one can find a certificate for that assumption, and reconstruct the
	state with explicit guarantees on the reconstruction error, from
	$O(c d \log^2d)$ Pauli expectation values. 
	The probability of failure is exponentially small in $c$.
\end{observation}

Finally, when the state is approximately low-rank but not nearly pure ($r>1$), one may perform tomography using different numbers of random Pauli expectation values $m$.  When $m$ is larger than necessary (corresponding to an over-estimate of $r$), we are guaranteed to find the correct density matrix.  When $m$ is too small, we find empirically that the algorithms for reconstructing the density matrix (i.e., solving the convex program (\ref{eqn:feasible})) simply fail to converge.  
 
\paragraph{A hybrid approach to matrix recovery.}  Here we describe a variant of our tomography method 
that makes the classical post-processing step (i.e., solving the convex program (\ref{eqn:feasible}) to 
reconstruct the density matrix) faster.  This method also uses random Pauli measurements, but they are chosen in a structured way.  
Any Pauli matrix is of the form $w(u,v) = \bigotimes_{k=1}^n i^{u_k
v_k} (\sigma^x)^{u_k} (\sigma^z)^{v_k}$ for $u,v\in\{0,1\}^n$.
We choose a
random subset $S \subset \set{0,1}^n$ of size $O(r \plog(d))$, and
then for all $u \in S$ and $v \in \set{0,1}^n$, measure the Pauli
matrix $w(u,v)$. 
We call this the ``hybrid method'' because it is equivalent to a
certain structured matrix completion problem.  This fact implies that
certain key computations in solving the convex program
(\ref{eqn:feasible}) can be implemented in time $O(d)$ rather than
$O(d^2)$ \cite{Cai2008}.  However, the hybrid method is not covered by
the strong theoretical guarantees shown earlier, though it does give
accurate results in practice.  For a more complete discussion, see the
EPAPS supplement.

\begin{figure}[t]
\begin{center}
	\includegraphics{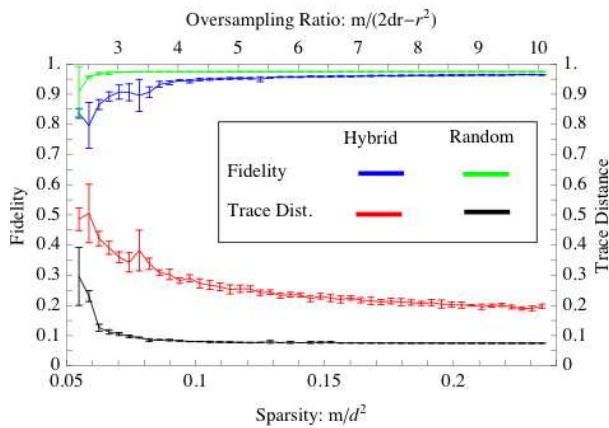}
	\caption{Average fidelity and trace distance vs.\ (scaled) number of measurement settings $m$ for random states of $n=8$ qubits, so $d=2^n$.  As discussed in the text, the sampled states had rank $r=3$, depolarizing noise of $5\%$ and Gaussian 
	statistical noise with $\sigma=0.1/d$.  Both the random Pauli and hybrid approaches are shown.}
	\label{F:fig1}
\end{center}
\vspace{-20pt}
\end{figure}

\paragraph{Numerical results.}
We numerically simulated both the random Pauli and hybrid approaches discussed above.  For both approaches, we used singular value thresholding (SVT)~\cite{Cai2008}.  Instead of directly solving Eq.~(\ref{E:cop}), 
SVT minimizes $\ \tau \| \sigma \|_{\tr} + \|\sigma\|_2^2/2 $ subject to
$|\tr(\sigma -\omega) w(A_i)| \le \delta$,
which is a good proxy to Eq.~(\ref{E:cop}) when $\tau$ dominates the 
second term; the programs are equivalent in the limit $\tau \rightarrow \infty$ (provided 
Eq.~(\ref{E:cop}) has a unique solution) \cite{Cai2008}.
Estimating the second term for typical states suggests choosing 
$2 \tau r \gg 1$; we use $\tau = 5$.  
To simulate tomography, we chose a random state from the Haar
measure on a $d \times r$ dimensional system and traced out the
$r$-dimensional ancilla, then applied depolarizing noise of strength
$\gamma$.  We sampled expectation values associated with randomly
chosen operators as above, and added additional statistical noise
(respecting Hermiticity) which was i.i.d.\  Gaussian with variance
$\sigma^2$ and mean zero.  We used SVT and
quantified the quality of the reconstruction by the fidelity and the
trace distance for various values of $m$, each averaged over $5$
simulations.  This dependence is shown in Fig.~\ref{F:fig1}.  The
reconstruction is remarkably high fidelity, despite severe
undersampling and corruption by both depolarizing and statistical
noise~\cite{tr_renorm}.  Using the hybrid method with $8$ qubits on a rank $3$ state plus
$\gamma = 5\%$ depolarizing, and statistical noise strength  $\sigma d
=0 .1$, we typically achieve $95\%$ fidelity reconstructions in under
$10$ seconds on a modest laptop with $2$ GB of RAM and a $2.2$ GHz
dual-core processor using MATLAB --- even though $90\%$ of the matrix
elements remain unsampled.  Increasing the number of samples only
improves our accuracy and speed, so long as sparsity is maintained.

Using truly randomly chosen Pauli observables (instead of the hybrid
method) slightly increases the processing time due to the dense matrix
multiplications involved: in our setup about one minute.  However,
this method achieves even better performance with respect to errors,
as seen in Fig.~\ref{F:fig1}.  

The simulations above show that our method work for generic low rank states.
Lastly, we demonstrate the functioning of the approach in the experimental
context of the state $\rho$ found in the $8$ ion experiment of 
Ref.\ \cite{Haffner2005}. To exemplify the above results,
we simulated physical measurements by sampling 
from the probability distribution computed using the Born rule applied to the 
reconstructed state $\rho$. This state is approximately low-rank, with 99\% of the weight concentrated on the first $11$ 
eigenvectors. The standard deviation per observable was $3/d$. Fewer than 30\% of 
all Pauli matrices were chosen randomly. From this information, a rank $= 3$ 
approximation $\sigma$ with fidelity of $90.5\%$ with respect to $\rho$ was found in about $3$ 
minutes on the aforementioned laptop.

\paragraph{Discussion.}  
We have presented new methods for low-rank quantum state tomography, 
which require only $O(r d \, \log^2(d))$ measurements, where $r$ is the rank of the 
unknown density matrix and $d$ is the Hilbert space dimension.  
Our methods are based on and further develop the new paradigm of 
compressed sensing, and in particular, matrix completion 
\cite{Candes2008,Candes2009a}.  We use measurements that are experimentally 
feasible, together with very fast classical post-processing.  
The methods perform well in practice, and are also supported by theoretical guarantees. 
It would be interesting to further flesh out the trade off between the 
need for measurements that can be performed easily in an experiment and 
 the need for sparse matrices during the classical post-processing step. It is the hope
 that this work stimulates such further investigations. 

{\it Acknowledgments.}
We thank E.\ Cand\`es and Y.\ Plan for useful discussions.  
Research at PI is supported by the Government of 
Canada through Industry Canada and by the Province of Ontario 
through the Ministry of Research~\& Innovation.  YL is 
supported by an NSF Mathematical Sciences Postdoctoral Fellowship,
JE by the EU (QAP, QESSENCE, MINOS, COMPAS) and the EURYI, 
DG by the EU (CORNER).  We thank the anonymous referees for many helpful suggestions.

\newpage

\section{Appendix}

\subsection{Details of the proof of Theorem 1}

While this publication contains a complete proof of all the claims
relevant for quantum tomography, the reader is invited to consult the
more general and explicit presentation in Ref.\ \cite{prep2} (and soon
\cite{prep1}). 
Below, we provide those details of the proof of Theorem~1, which were
left out in the main text.

We introduce some more formal notations used in the argument.  Denote
the trace inner product between two Hermitian operators $\rho, \sigma$
by $\bbraket{\rho}{\sigma}:=\tr \rho\sigma$.  We assume that $w(A_1),
\dots, w(A_m)$ are independent, identically distributed matrix-valued
random variables, with $w(A_i)$ drawn from the $d^2$ Pauli matrices
with uniform probability. Thus, we model the selection of the
observables as a process of sampling \emph{with} replacement. It is
both very plausible and easily provable \cite{nesme} 
that drawing the observables \emph{without}
replacement can only yield better results.

\subsubsection{Non-commutative large-deviation bound}

An essential tool for the proof is a non-commutative large-deviation
bound from \cite{ahlswede}. Let $S=\sum_i^m X_i$ be a sum of i.i.d.\
matrix-valued random variables (r.v.'s) $X_i$. Then it is shown in
\cite{ahlswede} that for every $\lambda, t>0$ we have 
\begin{equation}\label{se}
	\Pr[\|S\|>t] \leq 2d e^{-\lambda t} \bigl\|\EE[e^{\lambda X}]\bigr\|^m.
\end{equation}
It is simple to derive a Bernstein-type inequality from (\ref{se}).
Indeed, assume that $Y$ is some operator-valued random variable with
which is bounded in the sense that  $\|Y\|\leq1$ with probability one
and which has zero mean $\EE[Y]=0$. Recall the standard estimate
\begin{equation*}
	1+y \leq e^y \leq 1 + y + y^2 
\end{equation*}
valid for real numbers $y\in[-1,1]$ (actually a bit beyond). From the
upper bound, we get $e^Y \leq \Id + Y + Y^2$. From the lower bound:
\begin{eqnarray}
	&&\EE[e^Y] 
	\leq \Id + \EE[Y^2]
	\leq \exp(\EE[Y^2]) \nonumber \\
	&\Rightarrow&
	\|\EE[e^Y]\| \leq \|\exp(\EE[Y^2])\|=\exp(\|\EE[Y^2]\|).
	\label{eqn:ysquared}
\end{eqnarray}
In order to apply (\ref{eqn:ysquared}) to (\ref{se}), we set
$Y=\lambda X$. The parameter $\lambda$ is chosen to be
$\lambda={t}/{(2m\sigma^2)}$, where 
$\sigma^2 = \|\EE[X^2]\|$.
A straight-forward calculation now gives
\begin{eqnarray}\label{eqn:bernstein}
	\Pr[\|S\| > t]
	&\leq&
	2 d \, e^{-t^2/4 m \sigma^2},
\end{eqnarray}
(valid for $t\leq 2 m \sigma^2 / \|X\|$).

\subsubsection{``Case (i)'': large-deviation bound}

The first application of (\ref{eqn:bernstein}) is to verify
Eq.~(2) from the main text, which claims that
\begin{equation}\label{eqn:devNormEpaps}
	\Pr[\|\A-\Id_T\| > t]< 4dr e^{-t^2\kappa /8}.
\end{equation}

To this end, let $Y_i$ be the super-operator defined by
\begin{equation*}
	Y_i(\sigma)=\frac{d^2}{m} \P_T(w(A_i))\,
	\bbraket{w(A_i)}{\P_T(\sigma)}.
\end{equation*}
We will employ Eq.~(\ref{eqn:bernstein}) on the r.v.'s\ $X_i = (Y_i -
\EE[Y_i])$, where $\EE[Y_i]=\frac1m \Id_T$. 
From the fact that
$x\mapsto x^2$ is operator convex, one has $\sigma^2 = \|\EE[(Y-\EE
Y)^2]\| \leq \|\EE[Y^2]\|$. 
To estimate the latter quantity, we bound (using 
H\"older's inequality (c.f.\ [Bhatia, {\it Matrix Analysis}]))
\begin{eqnarray*}
	\|\P_T w_a\|_2^2 
	&=& \sup_{t\in T, \|t\|_2=1} (w_a, t)^2
	\leq \|w_a\|^2 \|t\|_{\tr}^2 \nonumber \\
	&\leq& \|w_a\|^2\, 2r \|t\|_2^2
	\leq 2 \frac rd.
\end{eqnarray*}
and hence
\begin{eqnarray*}
	\EE[Y^2]  
	&=& \frac{n^2}{m}
	\EE\big[(w_{A},\P_Tw_{A})\,Y\big]   \\
	&\leq& \frac{d^2}{m} \frac{2r}{d} 
	\EE\big[Y\big] = \frac{2d r}{m^2}\,\P_T. 
\end{eqnarray*}
which implies $\sigma^2\leq \frac{2dr}{m^2}$.
The claimed Eq.~(\ref{eqn:devNormEpaps}) directly follows by plugging this
estimate of $\sigma^2$ into the non-commutative large-deviation bound
(\ref{eqn:bernstein}).

\subsubsection{``Case (ii)'': the approximate subgradient}

Next, consider the claim after Eq.~(3) of the main text. There, we
assumed that $Y$ was a matrix in $\Span(w(A_1),\ldots,w(A_m))$ such
that
\begin{equation}\label{eqn:dualEpaps}
	\|\P_T Y - E\|_2 \leq 1/(2 d^2), \qquad \|\P_T^\bot Y \| < 1/2.
\end{equation}
It is to be shown that $\mathcal{R}\Delta=0$ implies
$\|\rho+\Delta\|_{\mathrm{tr}} > \|\rho\|_{\mathrm{\tr}}$.

Recall the scalar sign function $\sign$ which maps positive numbers to
$+1$, $0$ to $0$ and negative numbers to $-1$. If $\sigma$ is any
Hermitian matrix, then $\sign \sigma$ is the matrix resulting from
applying the $\sign$-function to the eigenvalues of $\sigma$. Note
that
\begin{equation}\label{eqn:trsign}
	\tr \sigma = (\sign\sigma,\sigma)
\end{equation}
and recall H\"older's inequality \cite{bhatia}
\begin{equation}\label{eqn:holder}
	(\sigma_1, \sigma_2) \leq \|\sigma_1\|_{\mathrm{tr}} \|\sigma_2\|
\end{equation}
for any two Hermitian $\sigma_1, \sigma_2$.

Letting $F=\sign\Delta_T^\perp$ we compute:
\begin{eqnarray}
	\|\rho+\Delta\|_{\tr} \nonumber
	&\geq&
	\|E(\rho+\Delta)E\|_{\tr}+
	\|(\Id-E)(\rho+\Delta)(\Id-E)\|_{\tr} \nonumber \\
	&\geq&
	(E,\rho+E \Delta E)+
	\big(F,\Delta_T^\bot\big) \nonumber \\
	&=&
	\|\rho\|_{\tr}+\big(E,\Delta_T\big)+\big(F,\Delta_T^\perp\big)-\big(Y,\Delta\big) \label{eqn:noisefree} \\
	&=&
	\|\rho\|_{\tr}+\big(E-\P_T Y,\Delta_T\big)+\big(F-\P_T^\bot Y,
	\Delta_T^\bot\big) \nonumber \\
	&>&
	\|\rho\|_{\tr} - \tfrac{1}{2d^2} \|\Delta_T\|_2 +\tfrac{1}{2}\|\Delta_T^\bot\|_{\tr} \geq \|\rho\|_{\tr}. \nonumber
\end{eqnarray}
(Use the ``pinching inequality'' \cite{bhatia} in the first step;
(\ref{eqn:trsign}), (\ref{eqn:holder}) in the second. The third step
is (\ref{eqn:trsign}) and using that $\mathcal{R}\Delta=0$ and
$Y\in\range Y$ implies $(Y,\Delta)=0$. The last estimate uses
(\ref{eqn:dualEpaps}) and, once more, (\ref{eqn:holder})).

\subsubsection{``Case (ii)'': large deviation bound}

The deviation bound before Eq.~(5) of the main text follows again from
(\ref{eqn:bernstein}). 
Let $F$ be an arbitrary matrix in $T$. 
With $X_i=\frac{d}{m} \P^\bot_T(w(A_i)) \tr w(A_i) F $:
\begin{eqnarray}
	\sigma^2
	&=&
	\sup_{\psi, \|\psi\|=1} \frac 1{d^2} \sum_a 
	\frac{d^2}{m^2} \big(\tr w_a F\big)^2 
	\bra\psi (\P_T^\bot w_a)^2 \ket \psi \nonumber\\
	&\leq&
	\frac{1}{m^2} \sum_a \big(\tr w_a F\big)^2 
	= \frac{d}{m^2} \|F\|_2^2, \label{eqn:sigmaBot}
\end{eqnarray}
having used that $\|\P_T^\bot w_a \|\leq 1$ and that the $\{d^{-1/2}
w_a\}$ form an orthonormal basis. 
Thus 
\begin{equation}\label{eqn:golfingEpaps}
	\Pr[\|\P^\bot_T\R F\| > t \|F\|_2] < 2d e^{-t^2 \kappa r/4}.
\end{equation}
In the proof, we use (\ref{eqn:golfingEpaps}) for $t=1/(4\sqrt r)$. Hence
the probability of failure becomes
\begin{equation*}
	p_3 \leq 2 d e^{-\frac{\kappa}{64}}.
\end{equation*}

\subsection{Details for Observation 1}

In this subsection we need to assume that the Paulis are sampled
\emph{without} replacement. All previous bounds continue to hold ---
see remark above.
Let 
\begin{equation*}
	\Q:\: \rho \mapsto \frac{1}{d} \sum_{i=1}^m w(A_i) \Tr \rho w(A_i)
\end{equation*}	
be the projection operator onto $\range \R$, normalized so that $\norm{\Q}=1$.  Define $\gamma = \frac{m}{d^2}$, and note that $\Q = \gamma\R$.  The optimization program (6) of the main text becomes $\min \|\sigma\|_{\tr}$, s.t. $\|\Q\sigma - \Q\omega\|_2 \leq \gamma\varepsilon$.  

Let $\Delta = \sigma-\rho$.  We upper-bound $\norm{\Q{\Delta}}_2$ as
follows.  First, 
\begin{equation*}
	\norm{\Q{\Delta}}_2 \leq \norm{\Q(\sigma-\omega)}_2
+ \norm{\Q(\omega-\rho_t)}_2 + \norm{\Q(\rho_t-\rho)}_2.  
\end{equation*}
For any
feasible $\sigma$, the first term is bounded by $\gamma\varepsilon$,
while the second term is bounded by $\gamma\varepsilon_2$.  For the
third term, note that for the fixed matrix $\rho_t-\rho$,
$\EE[\norm{\Q(\rho_t-\rho)}_2^2] = \gamma\norm{\rho_t-\rho}_2^2$, so
by Markov's inequality, $\norm{\Q(\rho_t-\rho)}_2^2 \leq \lambda^2
\gamma\norm{\rho_t-\rho}_2^2$, with probability at least
$1-\frac{1}{\lambda^2}$ \cite{markov}.
Thus we have 
\[
\norm{\Q{\Delta}}_2
 \leq \gamma\varepsilon + \gamma\varepsilon_2 + \lambda\sqrt{\gamma}\varepsilon_1
 \leq 2\gamma\varepsilon = 2\delta
\]
(where we defined $\delta = \gamma\varepsilon$).

On the other hand, we can also lower-bound $\norm{\Q{\Delta}}_2$ as follows:  $\|\Q\Delta\|_2 \geq \|\Q\Delta_T\|_2 - \|\Q\Delta_T^\bot\|_2$.  For the second term, we have $\|\Q\Delta_T^\bot\|_2 \leq \|\Delta_T^\bot\|_2$ (we cannot use Markov's inequality, because here we require a bound that holds simultaneously for all $\Delta$).  For the first term, recall from the noise-free case that $\A=\P_T\R\P_T$ satisfies $\|\Id_T - \A\|<1/2$ with high probability, and hence we have $\|\Q\Delta_T\|_2 \geq \gamma \|\A\Delta_T\|_2 \geq \tfrac{1}{2}\gamma \|\Delta_T\|_2$.  So we have 
\[
\|\Q\Delta\|_2 \geq \tfrac{1}{2}\gamma \|\Delta_T\|_2 - \|\Delta_T^\bot\|_2.
\]
Combining the above two inequalities and rearranging, we get 
\begin{equation}\label{eqn:noiseFeasible}
	\|\Delta_T\|_2 
	\leq \frac{2}{\gamma} \big( 2\delta + \|\Delta_T^\bot\|_2 \big)
	\leq \frac{2}{\gamma} \big( 2\delta + \|\Delta_T^\bot\|_{\tr} \big).
\end{equation}

We now show that $\|\rho+\Delta\|_{\tr}<\|\rho\|_{\tr}$ implies that $\Delta$ must be small.  With the estimate (\ref{eqn:noiseFeasible}) at our disposal, we re-visit (\ref{eqn:noisefree}):
\begin{eqnarray*}
	&&
	\|\rho+\Delta\|_{\tr}-\|\rho\|_{\tr}\\
	&\geq&
	 \big(E,\Delta_T\big) 
	+\big(F,\Delta_T^\perp\big) 
	-\big(Y,\Delta\big) 
	+\big(Y,\Delta\big) \\
	&=&
	 \big(E-\P_T(Y),\Delta_T\big) 
	+\big(F-\P_T^\perp(Y),\Delta_T^\perp\big) 
	+\big(Y,\Q(\Delta)\big) \\
	&>&
	-\frac{1}{2d^2} \|\Delta_T\|_2 + \frac{1}{2}\|\Delta_T^\bot\|_{\tr}
	-2\delta \|Y\|_2 \\
	&\geq&
	-\frac{1}{2d^2} \cdot \frac{2}{\gamma} (2\delta+\|\Delta_T^\bot\|_{\tr})
	+\frac{1}{2}\|\Delta_T^\bot\|_{\tr}
	-2\delta \|Y\|_2 \\
	&=&
	(\tfrac{1}{2} - \tfrac{1}{m}) \norm{\Delta_T^\perp}_{\tr} - 2\delta (\tfrac{1}{m} + \norm{Y}_2). 
\end{eqnarray*}
We use a crude bound $\norm{Y}_2 = \norm{\P_T(Y)}_2 + \norm{\P_T^\perp(Y)}_2 \leq \norm{\P_T(Y)-E}_2 + \norm{E}_2 + \norm{\P_T^\perp(Y)} \sqrt{d} \leq \frac{1}{2d^2} + \sqrt{r} + \frac{1}{2}\sqrt{d}$.  Then, for reasonable values of the parameters (say $d\geq 16$, $m\geq 16$, $r\leq d/10$), we have 
\[
\|\rho+\Delta\|_{\tr}-\|\rho\|_{\tr}
 > \tfrac{7}{16} \norm{\Delta_T^\perp}_{\tr} - 2\delta \sqrt{d}.
\]
So $\|\rho+\Delta\|_{\tr}<\|\rho\|_{\tr}$ implies 
\begin{equation} \label{eqn:noiseDeltaTPerp}
\norm{\Delta_T^\perp}_{\tr} < \tfrac{32}{7}\delta \sqrt{d}.
\end{equation}

Finally, write $\norm{\Delta}_{\tr} \leq \sqrt{2r} \norm{\Delta_T}_2 + \norm{\Delta_T^\perp}_{\tr}$, and use (\ref{eqn:noiseFeasible}) and (\ref{eqn:noiseDeltaTPerp}).  After simplifying, substituting in $\delta = \gamma\varepsilon$, and setting $\kappa = m/(rd)$, one obtains 
\begin{equation}
\begin{split}
\norm{\Delta}_{\tr}
 &\leq 6\varepsilon\sqrt{r} + 13\varepsilon\sqrt{rd} + 5\varepsilon\frac{\kappa r}{\sqrt{d}}
  \leq O(\varepsilon \sqrt{rd}).
\end{split}
\end{equation}

Finally, we write $\norm{\sigma-\rho_t}_{\tr} \leq \norm{\sigma-\rho}_{\tr} + \norm{\rho-\rho_t}_{\tr}$.  The first term is bounded by $O(\varepsilon \sqrt{rd})$ as shown above; the second term is $\leq \norm{\rho-\rho_t}_2 \sqrt{d} \leq \varepsilon_1 \sqrt{d} \leq \varepsilon \sqrt{d}$.  This gives the desired result.

\subsection{Certified tomography for almost-pure states}

For almost-pure states ($r=1$), it is
possible to obtain estimates for $\delta_1$ from only
$O(d)$ Pauli expectation values without any
assumptions. In this subsection, we sketch a simple scheme based on this
observation: it outputs a reconstructed density matrix $\sigma$,
together with a certified bound on the deviation
$\|\sigma-\rho_t\|_{\tr}$. The algorithm takes two inputs: $O(d\,\log^2d)$
random Pauli expectation values, and the experimentalist's estimate of
the measurement precision $\delta_2$ \cite{errormodel}.  

Concretely, we set $r=1$ and aim to put a bound on
$\delta_1=\|\rho_t-\ket\psi\bra\psi\|_2$, where $\ket\psi$ is the
eigenvector of $\rho_t$ corresponding to the largest eigenvalue.  Such
a bound can be obtained in terms of the \emph{purity} $\tr
\rho_t^2=\|\rho_t\|_2^2$. E.g.,\
\begin{equation}
	\delta_1 = \| \rho_t - |\psi\rangle\langle\psi| \|_2 \leq 
	2^{1/2} (1-\|\rho_t\|_2^2) \label{eqn:purity}
\end{equation} 
(valid for $\|\rho_t\|\geq 1/2$, which can certifiably be tested).
Estimating the purity is done in a way analogous to the proof of
Theorem~1. Choose $m$ i.i.d.\ random variables $A_i$ taking values in
$[1,d^2]$,
and define  $S=(d/m) \sum_{i=1}^m |\tr w(A_i) \omega|^2$. Then
$\EE[S]=\|\omega\|_2^2$ and thus $\|\rho_t\|_2 \geq
\EE[S]^{1/2}-\delta_2$.  
We can bound the
deviation of $S$ from its expected value by the standard (commutative)
Chernoff bound. One finds for the variance
$
	\operatorname{Var}((d/m)|\tr w(A) \omega|^2) \leq (d/m^2)
	\|\omega\|_2^2 \leq d/m^2,
$
so that (for $t\in[0,1]$):
\begin{eqnarray*}
	 \Pr\bigl[\bigl|S - \|\omega\|_2^2\bigr|>t \bigr]
	 &\leq& 2 e^{-t^2 m/(4d)},
\end{eqnarray*}
Choose $m=4\mu d/t^2$ for some $\mu>1$ to ensure that
\begin{eqnarray}
 	\Pr[  |S - \|\rho_t\|^2_2 |> t + 2 \delta_2 +\delta_2 ^2] < e^{-\mu}.
\end{eqnarray}
Combining the previous equation with (\ref{eqn:purity}), we have
arrived at a certified estimate for $\delta_1$. \\

\subsection{A hybrid approach to matrix recovery}

Matrix recovery using Pauli measurements does lack one desirable feature:  the
classical post-processing (solving the convex programs) is more
costly, compared to matrix completion \cite{Candes2008,Candes2009a}.
This is due to the role of sparse linear algebra in the SVT (singular
value thresholding) algorithm \cite{Cai2008}.  
The basic issue is that SVT must handle matrices of the form
$\R\rho$.  For matrix completion, $\R\rho$ is 
sparse, so basic operations such as matrix-vector multiplication take
time $O(d)$; but when we use random Pauli measurements,
$\mathcal{R}(\rho)$ is dense, and basic operations take time $O(d^2)$.  
We now describe a ``hybrid'' approach that avoids this difficulty, and
works well in practice. The main observation is that for certain,
carefully selected sets of Pauli matrices, $\R\rho$ is sparse after
all. 

Any Pauli matrix is of the form 
$$
	w(u,v) = \bigotimes_{k=1}^n i^{u_k v_k} (\sigma^x)^{u_k} (\sigma^z)^{v_k}
$$
for $u,v\in\{0,1\}^n$. Plainly, the position of the $d$ non-zero matrix elements 
of $w(u,v)$ depends only on $u$
($v$ encodes only phase information).  Now choose a
random subset $S \subset \set{0,1}^n$ of size $O(r \plog(d))$, and
then for all $u \in S$ and $v \in \set{0,1}^n$, measure the Pauli
matrix $w(u,v)$. Thus we are measuring each of the Pauli strings containing only $\sigma^z$ or identity, together with these same strings ``masked'' by applying a set of size $|S|$ of Pauli strings with a pattern of $\sigma^x$ and identity.  Formally, this means 
$$
	\R\rho\propto \sum_{u\in S, v \in\{0,1\}^n} w(u,v) \tr(\rho w(u,v)).
$$
It follows that $\R\rho$ is sparse with only $|S|d$
non-zero matrix elements.  This ``hybrid method'' can be viewed as a variant of the usual matrix completion problem, where instead of sampling matrix elements independently at random, we sample groups of matrix elements determined by the random strings $u \in S$.

While the hybrid algorithm works well for generic states, certain
input states $\rho$ may fail to be ``incoherent enough'' w.r.t.\ the
very specific set expectation values obtained (c.f.\
\cite{Candes2008,Candes2009a}).  For example, when the eigenvectors of
$\rho$ are nearly aligned with the standard basis, most of the matrix
elements of $\rho$ are nearly 0, and hence matrix completion is
impossible. To avoid this problem, we suggest to perform a
pseudo-random unitary $U$ prior to measuring the Pauli matrices. One
then uses the hybrid method on $U\rho U^{\dagger}$, and finally
applies $U^{-1}$ to recover $\rho$. In particular, one can draw $U$
at random from an (approximate) unitary $k$-design with $k \sim n/\log
n$.  Explicit constructions of such unitaries are known, and can be
implemented efficiently \cite{harrow-low}.  

While we cannot at this point prove rigorous guarantees for the hybrid
approach, we do show below that randomization by approximate
$k$-designs generates sufficient ``incoherence'' that the original
matrix completion algorithms \cite{Candes2008,Candes2009a} would work.
Because these algorithms call for matrix elements to be sampled from a
uniform distribution, Observation~\ref{obs-incoherence} does not
rigorously apply to the hybrid scheme. It does, however, make it
\emph{plausible} that pseudo-randomization overcomes incoherence
problems and that guarantees for the hybrid method can be proven in
the future.

\begin{observation}[Incoherence from $k$-designs] \label{obs-incoherence}
Let $\rho$ be an arbitrary state of rank $r$ and dimension $d$, and
let $E$ be the projector onto the support of $\rho$.  Let $\ket{i}$,
$i=1,\ldots,d$, denote the standard basis.  Let $U$ be drawn at random
from an ($\varepsilon$-approximate) unitary $k$-design with $k \sim
n/\log n$ (and $\varepsilon = 1/d^k$), and let $\ket{b_i} = U\ket{i}$.
Then, with probability at least $1-(1/d)$, the following holds:
\[
\text{for all $i=1,\ldots,d$, } \norm{E\ket{b_i}}^2_2 \leq \mu_0 r/d, 
\]
where $\mu_0 = C_1(\log d)^{C_2}$, and $C_1$ and $C_2$ are fixed
constants.
\end{observation}

This implies the incoherence conditions (A0) and (A1) of
\cite{Candes2008}, specialized to the case of positive semidefinite
matrices, with $\mu_0$ as given above and $\mu_1 = \mu_0\sqrt{r}$.
Combining with the results of \cite{Candes2008} shows that ordinary
matrix completion, with matrix elements sampled independently at
random, will succeed.  This guarantee does not extend to the hybrid
method, however.

Proof of Observation \ref{obs-incoherence}:  First consider a single
vector $\ket{b_1}$, and define $Z = \norm{E\ket{b_1}}^2_2$.  We will compute the $k$'th moment of $Z$:
\begin{equation*}
\begin{split}
\EE[Z^k] &= \EE[\Tr(E^{\otimes k} \ket{b_1}\bra{b_1}^{\otimes k} E^{\otimes k})] \\
 &= \Tr(E^{\otimes k} \EE[\ket{b_1}\bra{b_1}^{\otimes k}] E^{\otimes k}).
\end{split}
\end{equation*}

We want to compute $\EE[\ket{b_1}\bra{b_1}^{\otimes k}]$.  Let $\ket{u_1}$ be a Haar-random unit vector in $\CC^d$, and let 
\[
\Delta = \EE[\ket{b_1}\bra{b_1}^{\otimes k}] - \EE[\ket{u_1}\bra{u_1}^{\otimes k}].  
\]
By the definition of an approximate unitary $k$-design, every matrix
element of $\Delta$ has absolute value at most $\varepsilon/d^k$.
Thus $\norm{\Delta}_2 \leq \varepsilon$.  
A well-known (c.f.\ e.g.\ Def.\ 2.1 in \cite{harrow-low-cmp})
corollary of Schur's Lemma states
$\EE[\ket{u_1}\bra{u_1}^{\otimes k}] = \Pi_S/\dim(S)$, where $S$ is
the symmetric subspace of $(\CC^d)^{\otimes k}$, $\Pi_S$ is the
projector onto $S$, and $\dim(S) = \binom{d+k-1}{k}$.  So we have 
\[
\EE[\ket{b_1}\bra{b_1}^{\otimes k}] = \frac{\Pi_S}{\dim(S)} + \Delta.
\]

Substituting in, we get:
\[
\begin{split}
\EE[Z^k] &= \frac{\Tr E^{\otimes k} \Pi_S}{\dim(S)} + \Tr E^{\otimes k} \Delta \\
 &\leq \frac{\norm{E^{\otimes k}}_{\tr} \norm{\Pi_S}}{\dim(S)} + \norm{E^{\otimes k}}_2 \norm{\Delta}_2 \\
 &\leq \frac{r^k k!}{(d+k-1)\cdots d} + \varepsilon\sqrt{r^k} \leq \Bigl(\frac{rk}{d}\Bigr)^k.
\end{split}
\]
Using Markov's inequality, and setting $t = (rk/d)\cdot d^{2/k} \leq (r/d)\cdot \text{poly}(\log d)$, we get 
\[
\Pr[Z>t] \leq \frac{\EE[Z^k]}{t^k} \leq \Bigl(\frac{rk}{td}\Bigr)^k = \frac{1}{d^2}.
\]
This proves the claim for a single vector $\ket{b_1}$.  Now take the union bound over all the vectors $\ket{b_i}$, $i=1,\ldots,d$.  $\square$


\begin{thebibliography}{10}
\bibitem{note2} 	
	We use the usual matrix norms $\|A\|_{\tr}=\sum_i
	\sigma_i, \|A\|_2^2=\tr A^\dagger A = \sum_i \sigma_i^2, \|A\|=\max_i
	\sigma_i$, with $\sigma_i$ the singular values of $A$. The last
	definition extends to super-operators: if $\mathcal{A}$ is a
	super-operator, then $\|\mathcal{A}\|$ is its largest singular value,
	or, equivalently $\|\mathcal{A}\|=\sup_{\sigma, \|\sigma\|_2 = 1}
	\|\mathcal{A}\sigma\|_2$ (a.k.a.\ ``$2\to 2$''-norm).
\bibitem{note3} 
	If the term $1/(2d^2)$ were zero, $Y$ would be a
	strict subgradient.
\bibitem{localnoise}
	Consider a pure state of $n$ qubits subject to local noise that occurs with probability $p$ on each site.  Then the density matrix is well-approximated by a matrix of rank $r = 2^{nH(p)} = d^{H(p)}$, 
	where $H(p)$ is the binary entropy of $p$, and $d=2^n$ is the Hilbert 	
	space dimension.  When $p$ is small, we have $r \ll d$.
\bibitem{markov}
	Using Theorem~11 of \cite{prep2}, the Markov bound can immediately
	be replace by a more sophisticated large deviation bound, which
	gives a probability of failure exponentially small in $\lambda$.
\end{thebibliography}
\end{document}